\documentclass[11pt,a4paper]{article}
\pdfoutput=1
\usepackage{jheppub}
\usepackage{amsfonts,mathtools,graphics,slashed}

\usepackage{calligra} \DeclareMathAlphabet{\mathcalligra}{T1}{calligra}{m}{n} \DeclareFontShape{T1}{calligra}{m}{n}{<->s*[2.2]callig15}{}
\newcommand{\sr}{\ensuremath{\mathcalligra{r}}}

\DeclareMathOperator{\Tr}{Tr}
\newcommand{\dd}{\mathrm{d}}

\title{``\boldmath{$1k_F$}" Singularities and  Finite Density ABJM Theory
 at Strong Coupling}

\author[1]{ Oscar Henriksson}
\author[2]{Christopher Rosen}

\affiliation[1]{Department of Physics\\and Center for the Theory of Quantum Matter\\ 
 University of Colorado\\ Boulder, CO 80309, U.S.A}
 \affiliation[2]{Blackett Laboratory, 
  Imperial College\\ London, SW7 2AZ, U.K.}

\emailAdd{oscar.henriksson@colorado.edu}
\emailAdd{c.rosen@imperial.ac.uk}

\preprint{Imperial/TP/2016/CR/03 \,\, COLO-HEP-592}
\arxivnumber{
}

\abstract{
We study non-analytic behavior in the static charge susceptibility in finite density states of the ABJM theory using its holographic dual. Emphasis is placed on a particular state characterized by vanishing entropy density at zero temperature, and Fermi surface-like singularities in various fermionic correlation functions. The susceptibility exhibits branch points in the complex momentum plane, with a real part quantitatively very similar to the location of the Fermi surface singularities. 
}

\begin{document}
\maketitle
\flushbottom

\section{Overview}

Within the Landau Fermi liquid paradigm, the existence of a Fermi surface in a finite density phase of interacting matter is reflected in the analytic properties of various susceptibilities. More specifically, in conventional Fermi liquids a Fermi surface at momentum $k_F$ appears as a non-analyticity in (for example)  the static charge susceptibility $\chi(k)$ at $k=2k_F$.  That such a relationship between fermionic and bosonic response should exist is not surprising. In a Fermi liquid, charge transport is the responsibility of the excitations very near the Fermi surface, with momentum $k = k_F+q$ where $q/k_F \ll 1$. It is thus natural that the static susceptibility, which quantifies the response of the liquid to a charged impurity, encodes the length scale $1/k_F$ characteristic of these low lying excitations.

These so-called ``$2k_F$" singularities in susceptibilities arise on very general grounds, essentially as a consequence of kinematic constraints relating parallel portions of the Fermi surface. They are also fundamentally quantum mechanical in origin, owing to the fact that fermionic excitations near the Fermi surface have Compton wavelength $\lambda \sim 1/k_F$. The effects of this finite size on charge screening can already be anticipated from elementary quantum scattering considerations. A familiar textbook example (see e.g. \cite{AltSim,pCol} for this and related review below) points out that a free quantum mechanical particle (``electron") with momentum $k_F$  incident on a potential (``charged impurity") will give rise to a deviation in the charge density like
\begin{equation}
\delta|\Psi(x)|^2  = \Big|e^{ik_F x}+|\mathcal{R}|\,e^{-i(k_Fx-\delta)} \Big|^2-\Big|e^{ik_F x}\Big|^2= 2|\mathcal{R}|\cos(2k_Fx-\delta) +\mathcal{O}(|\mathcal{R}|^2)
\end{equation}
to the left of the impurity at the origin. These oscillations in the induced charge density, due the quantum mechanical ``size" of the relevant excitations, are an example of Friedel oscillations. Evidently, they are characterized by a wave vector of magnitude $2k_F$.

The quantum nature of these oscillations is also borne out by more explicit calculation. In the ubiquitous Random Phase Approximation (RPA), the dominant contribution to the susceptibility is captured by the particle--hole polarization diagram---a one loop effect. In 2+1 dimensions, computation of this diagram leads to the well known Lindhard function and static susceptibility of the form
\begin{equation}
\chi(k) \propto 1-\Theta\left(\frac{k}{k_F}-2\right)\sqrt{1-\left(\frac{2k_F}{k}\right)^2}.
\end{equation}
Written in this way, the $2k_F$ singularity is easily seen to coincide with the branch points of the square root. In the linear response regime, this non-analyticity in $\chi$ will leave its mark on the induced charge density $n$, giving rise to both oscillations and a characteristic power law decay:
\begin{equation}
n(\mathfrak{R}) \approx \frac{\cos\left(2k_F\,\mathfrak{R}+\delta\right)}{\mathfrak{R}^2}.
\end{equation}
Here $\mathfrak{R}$ gives the distance from the impurity (see also the discussion around (\ref{eq:nFrie})  below) and $\delta$ is again a phase. This expression makes only very mild assumptions about the form of the impurity potential, and thus retains its validity in a great variety of physical systems.

In particular, its realm of applicability is broadly expected to extend beyond Fermi liquids to include any system  with a ``well-defined" Fermi surface. An operational definition of the latter could be a sharp surface in momentum space such that gapless fermionic degrees of freedom congregate at this $k=k_F$ at zero temperature. In this case wave vectors connecting parallel patches of the Fermi surface ought to dominate the correlation functions, giving rise to $2k_F$ singularities in the susceptibility. Not surprisingly, comparitively little is known about the fate of the Friedel oscillations and other $2k_F$ singularities at strong coupling, or in systems that fall outside the paradigm of Landau Fermi liquidity. Attempts to quantify the effects of interactions on both the Lindhard function as well as the signature of Friedel oscillations have appeared in a variety of works, relying on a variety of theoretical frameworks (see \cite{2016PhRvB..93t5117D,PhysRevB.72.045127,2010PhRvL.104w6401A,Nayak:1994ng,Mross:2010rd,
PhysRevB.50.14048,2008PhRvB..78c5103S, Fitzpatrick:2014cfa} for a far from exhaustive sample). Although the specifics vary, the take home message seems to be that while interactions may ``soften" the $2k_F$ singularities, they typically remain in the response.

Various attempts to clarify  the situation have also been made within the framework of gauge/gravity duality \cite{Puletti:2011pr,Faulkner:2012gt,Polchinski:2012nh, Goykhman:2012vy, Hartnoll:2012wm,Anantua:2012nj, Blake:2014lva, Gouteraux:2016arz}. This approach is particularly attractive, in that it allows simple computational access to finite density phases of strongly interacting matter. In practice, no assumption of adiabaticity need apply with respect to the interactions, and thus the reach of this method extends beyond many conventional techniques. What's more, if the calculations are performed within a supergravity (SUGRA) theory that descends from a low energy limit of string or M theory, there exists the possibility of performing a careful match between the gravitational physics and properties of a known gauge theory in appropriate limits. 

In an early attempt to identify an analogue of Friedel oscillations in holographic matter, the authors of \cite{Puletti:2011pr} forgo the ``top-down" string/M-theory pedigree and study the polarization function in a semi-classical bulk model in which fermionic modes fill out a {\it bulk} Fermi sea. Although they succeed in constructing a model with a boundary Lindhard function that displays non-analytic behavior at $q=2k_F$, in this work we would like to adopt a slightly different perspective of holographic matter at finite density. 

In explicitly known examples of $AdS$/CFT duality, the holographic (boundary) gauge theories generically describe interacting bosons and fermions charged under various global symmetries. This is typically a consequence of large amounts of supersymmetry, in which case the global symmetries may be $R$-symmetries, for example. By turning on a chemical potential for matter charged under (typically the Cartan of) this symmetry, one effects a finite density deformation of the original theory. Generic states in the finite density theory are thus characterized by a large number of strongly interacting bosonic and fermionic modes, at least in the classical SUGRA limit of the duality. 

By adopting this perspective, our tastes seem to be more closely aligned with those of \cite{Blake:2014lva}, who search for charge density oscillations in the quintessential finite density phase of holographic matter---the $AdS$--Reissner-Nordstr\"om ($AdS$--RN) solution. Under standard application of the holographic dictionary, the bulk $U(1)$ gauge field maps to a conserved global current in the boundary theory. What exactly that global current is depends on the details of the holographic realization.  If the solution comes from a sector of the maximal gauged SUGRA in four dimensions, for example, the conserved current might be constructed from ABJM \cite{Aharony:2008ug} matter charged under a particular $U(1)$ subgroup of the $SO(8)_R$ symmetry. In what follows we will be primarily interested in interpreting our results holographically within this maximal gauged SUGRA/ABJM correspondence. 

Considered from this vantage point, the finite density ABJM phase dual to the $AdS$--RN solution is a strongly interacting plasma of bosonic and fermionic degress of freedom charged under the corresponding $U(1)$. Very generally, it is natural to wonder if the finite density of fermions present in this phase have arranged themselves into a Fermi circle in the 2+1 dimensional field theory. If so, then one would expect to find some indication of the length scale $1/k_F$ in various correlation functions, such as the static susceptibility as highlighted above. 

In \cite{Blake:2014lva} the authors compute this static susceptibility holographically, and find that there is in fact non-analytic behavior tied to one such length scale. Consequently, they observe oscillations in the induced charge density with wavelength controlled by this scale, providing an example of a strongly interacting analogue of the familiar Friedel oscillations. Very interestingly, the authors were able to understand this length scale analytically. They observe that the oscillations are controlled by branch points in a certain scaling exponent $\nu_{k}^-$ that governs the low energy fluctuations of the charge density.

Explicitly, this exponent calculated in the $AdS$--RN background turns out to be given by
\begin{equation}\label{eq:nu_kAdS}
\nu_k^- = \frac{1}{2}\sqrt{5+8\left(\frac{k}{\mu}\right)^2-4\sqrt{1+4\left(\frac{k}{\mu}\right)^2}}
\end{equation}
and the branch points that appear to be responsible for the induced oscillations are those at $k^\star /\mu=1/2\sqrt{2}+i/2$ where $\mu$ is the chemical potential. For future reference, we note that $\mathrm{Re}\,k^\star\approx 0.35\mu$ sets the scale on which the charge density oscillations take place.

Thus, the situation seems to be that the static susceptibility computed in this phase behaves as though there are charged degrees of freedom with a characteristic momentum $\sim\mathrm{Re}\,k^\star/\mu$. The obvious question, then, is whether or not there is  good reason to associate this characteristic momentum with the existence of a Fermi surface at $2k_F= \mathrm{Re}\,k^\star$ in the finite density phase. 

A sensible approach towards an answer to this question is to look for hints of a Fermi surface in other correlation functions evaluated in the same holographic phase. A particularly good candidate might be retarded two point functions of various fermionic operators in the ABJM theory. At zero temperature, such two point functions can be used to diagnose both the existence of a Fermi surface, where
\begin{equation}\label{eq:FSrule}
G^{-1}_R(\omega=0,k=k_F) = 0
\end{equation} 
as well as to understand the nature of fermionic excitations near the Fermi surface (via the spectral function, proportional to $\mathrm{Im}\,G_R$). 

By fully decoupling 32 of the 56 spin 1/2 modes of the maximal gauged SUGRA from all other fermions, the authors of \cite{DeWolfe:2011aa, DeWolfe:2014ifa} were able to compute such fermionic two-point functions in the ABJM phase dual to $AdS$--RN holographically. Two important lessons from that work are: 
\begin{enumerate}
\item Two-point functions of fermionic ABJM operators can exhibit Fermi surface like singularities at $\omega = 0$ and $k = k_F$ in the finite density phase dual to $AdS$--RN.
\item For the modes considered, these Fermi surface singularities arise at $k_F\approx 0.37\mu$.
\end{enumerate}

The numerical similarity between $\mathrm{Re}\, k^\star$ computed from the charge susceptibility and $k_F$ computed from fermion response---two logically independent calculations---is surprising. This similarity has hitherto gone unnoticed in the published literature, and forms the primary motivation for the present work.   

It is perhaps most surprising when viewed in the context of \cite{Huijse:2011ef,Iqbal:2011bf}, the results of which have encouraged an association between such holographic Fermi surfaces and carriers of ``cohesive" charge, whose density is supposed to be sub-leading as the number of colors in the gauge theory $N$ becomes large. In such an interpretation, there is no obvious reason to expect signatures of the linearized fermion response to manifest in the static susceptibility of the background charge density. Alternatively, \cite{DeWolfe:2011aa,Cosnier-Horeau:2014qya} have used elements of the holographic dictionary provided by a top-down embedding to suggest that one might indeed anticipate a connection between holographic boson and fermion response at leading order in large $N$.  A useful review of both of these ideas can be found in \cite{Hartnoll:2016apf}. It is also surprising that the holographic results appear to suggest the presence of  ``$1k_F$" singularities in the charge response, as opposed to the omnipresent $2k_F$ singularities that typically accompany a sharp Fermi surface as outlined above.

 In what follows, the goal will be to further explore the connection between non-analyticities in the static susceptibility and poles in fermionic two-point functions in finite density phases of ABJM matter. Ultimately, we will find some indication that such a connection may persist even in more complicated phases of holographic matter. This evidence will in turn depend strongly on our adherence to a ``top-down" framework for our holographic calculations, in which all bulk couplings for bosonic and fermionic fluctuations are fully determined by the maximal gauged SUGRA.

From the perspective of the fermion response, the holographic ABJM phase dual to $AdS$--RN is quite unlike a conventional Fermi liquid. Although there is a sharp Fermi surface in the sense of (\ref{eq:FSrule}), it was found in \cite{DeWolfe:2011aa} that the low lying fermionic degrees of freedom can not be interpreted as well defined Landau quasi-particles. Such a peculiarity could conceivably muddle any connection between ``$1k_F$" singularities in the fermion and charge density correlation functions. In an attempt to circumvent this potential obstruction, in this work we will focus on a holographic phase of matter that more closely resembles a conventional Fermi liquid from the perspective of top-down fermion response.

Towards this end, we will set the stage for our calculation in section \ref{setup} where we introduce the particular solution of $\mathcal{N}=8$, $D=4$ gauged SUGRA we wish to study. Interpreted holographically, this solution provides a gravitational description of a curious phase of ABJM matter. In particular, the phase is characterized by a specific heat that vanishes linearly at low temperatures, as well as by fermionic spectral functions which suggest the presence of stable charged excitations which reside near a Fermi surface (reviewed in section \ref{sec:num_construction}). Taken together, these features suggest that in such a phase the physics could in some sense be more ``Fermi liquid-like" and the connection between charge and fermion response might be even more pronounced. It will turn out that this does not seem to be the case.

One reason will be intimately related to the appearance of new channels of charged response in this background relative to the $AdS$--RN background.  The IR portion of this solution and its fluctuations has previously starred in related holographic investigations of current-current correlators as an ``$\eta$--geometry", with $\eta = 1$ \cite{Hartnoll:2012wm,Anantua:2012nj}. There the authors also arrive at the conclusion that such geometries are dual to phases that are in some sense more ``fermionic", albeit for rather different reasons. The fluctuation analysis in this background is discussed in section \ref{sec:statsus} (and also in more detail in \cite{Anantua:2012nj}) with some finer points relegated to an appendix.

 In contrast to the analysis performed in \cite{Anantua:2012nj}, the computation of the static susceptibility is sensitive to the entirety of the bulk geometry. There is thus little hope of solving exactly the coupled fluctuation equations, and to make progress we resort to numerical integration of these equations throughout the complex momentum plane. The results of this integration are shown in a series of plots in section \ref{sec:rez}, which provide the primary output of 
this work. 

We then turn to a discussion of our results in section \ref{disc}, where we highlight the underlying structure responsible for the analytic properties of the static susceptibility, and attempt to reconcile those properties with the fermion response.  Here we comment on the somewhat surprising similarities between Fermi momenta and branch points of the static susceptibility, and conclude with some avenues for further study.

\section{The 3--Charge Black Brane}\label{setup}
As suggested above, our interest lies primarily in the properties of finite density states of strongly interacting ABJM matter at zero temperature. The holographic dictionary thus directs us towards solutions of $D=4$ $\mathcal{N}=8$ gauged supergravity that asymptote to the maximally symmetric $AdS_4$ in the UV, but are driven elsewhere in the IR by a non-vanishing profile for an Abelian gauge field. 

Many such solutions are known to exist. To construct them, it is often convenient to consider a truncated subset of the gauged SUGRA which contains fewer fields. In this work, we will focus our attention on the properties of a particular solution which resides in the truncation of the SUGRA to singlets under the Cartan $U(1)^4 \subset SO(8)$ of the gauge symmetry. The relevant truncation (as well as its electrically charged black brane solutions) was worked out in \cite{Cvetic:1999xp}, and there the truncation's consistency was demonstrated by explicitly providing the lift to the eleven dimensional theory.

The truncation to the Cartan leaves an $\mathcal{N}=2$ SUGRA coupled to three vector multiplets. The so-called 3--charge black brane solution is a background in which three of the Abelian gauge fields are set equal to one another, while the fourth vanishes. This restriction also identifies the three dilatons with each other. Additionally, for the purely electric solution we consider, the axions all vanish. 

This sector of the SUGRA is thus described by a simple Lagrangian of the form
\begin{equation}\label{eq:Lag}
e^{-1}\mathcal{L} = R-\frac{1}{2}\left(\partial\phi\right)^2-\frac{3}{4}e^{\frac{\phi}{\sqrt{3}}}F^2+6\cosh \frac{\phi}{\sqrt{3}}
\end{equation}
where $\phi$ is the remaining dilaton and $F=\dd A$ is the field strength for the active $U(1)$ gauge fields. We work in units such that the maximally symmetric $AdS_4$ has radius $L= 2\kappa^2 = 1$. The electric 3--charge black brane solution is conveniently written in terms of the ansatz
\begin{align}\label{eq:bkg}
\mathrm{d}\bar{s}^2= \bar{g}_{\mu\nu}\dd x^\mu\dd x^\nu=e^{2\chi(r)}\Big(-f(r)\mathrm{d}t^2+\mathrm{d}\vec{x}^2\Big) +\frac{e^{-2\chi(r)}}{f(r)}\mathrm{d}r^2,\qquad \bar{A} = \Phi(r)\mathrm{d}t,\nonumber\\\mathrm{and} \qquad \bar{\phi} = \bar{\phi}(r).
\end{align}
Bars have been added to some quantities in anticipation of our impending fluctuation analysis, in which the bar will be used to denote background quantities. The solution is then given by
\begin{align}
\chi(r)&= \log r +\frac{3}{4}\log\left(1+\frac{Q}{r}\right) & f(r) &= 1-\left(\frac{r_H+Q}{r+Q} \right)^3\nonumber\\
\Phi(r) & = \sqrt{Q(r_H+Q)}\left(1-\frac{r_H+Q}{r+Q}\right) & \bar{\phi}(r) & = \frac{\sqrt{3}}{2}\log\left(1+\frac{Q}{r}\right)
\end{align}
where $r_H$ and $Q$ are integration constants that parametrize the location of the horizon and (roughly) the brane's charge. 

\subsection{Thermodynamics and Field Theory Interpretation}
Under the standard holographic interpretation of this solution, it is easy to see that the dual state of the ABJM theory is characterized by a temperature and entropy
\begin{equation}
T = \frac{3}{4\pi}\sqrt{r_H(Q+r_H)}\qquad \mathrm{and} \qquad s = 4\pi\sqrt{r_H}\left(r_H+Q\right)^{3/2}
\end{equation}
respectively. The solution contains bulk gauge and matter fields, and their role from the ABJM theory perspective is most conveniently understood from the near boundary behavior of these fields in Fefferman-Graham coordinates. In these coordinates the  near-boundary metric locally takes the form
\begin{equation}
\mathrm{d}s^2 = \sr^2g_{ij}\mathrm{d}x^i\mathrm{d}x^j+\frac{\mathrm{d}\sr^2}{\sr^2},
\end{equation}
while the gauge field and scalar fall off like
\begin{align}
\Phi(\sr\to\infty) &\approx \mu-\frac{\rho}{\sr}+\ldots\\
\bar{\phi}(\sr\to\infty) & \approx  \frac{\sqrt{3}Q}{2\sr}+\frac{\sqrt{3}Q^2}{8\sr^2}+\ldots .\label{eq:bkgSFO}
\end{align}
The coefficients governing the fall-offs of the gauge field have been labeled suggestively. They are interpreted holographically as a deformation of the ABJM theory by a non-vanishing chemical potential ($\mu$), which in turn places the dual theory in a finite density ($\rho$) phase. Explicitly, 
\begin{equation}
\mu = \sqrt{Q(r_H+Q)} \qquad \mathrm{and} \qquad \rho = (Q+r_H) \sqrt{Q(r_H+Q)}.
\end{equation}

The thermodynamic quantities are parametrized by two dimensionful parameters, $r_H$ and $Q$. With an eye towards the grand canonical ensemble, it is helpful to trade these parameters for $T$ and $\mu$, which allows us to express the charge and entropy densities as
\begin{equation}
\rho = \frac{\mu}{3}\sqrt{16\pi^2T^2+9\mu^2} \qquad \mathrm{and} \qquad s = \frac{16}{9}\pi^2 T \sqrt{16\pi^2T^2+9\mu^2}.
\end{equation}

The correct interpretation of the boundary behavior of the bulk scalar is a bit more subtle. If we denote the leading $1/\sr$ fall-off by $\alpha$ and the subleading $1/\sr^2$ fall-off by $\beta$, a naive application of the holographic dictionary would relate $\alpha$  to a deformation of the ABJM theory by a scalar operator, and $\beta$  to that operator's response. These are Dirichlet ($\alpha=0$) boundary conditions. It is long known, however, that in order to respect the supersymmetry of the gauged SUGRA these scalars must obey an ``alternate" ($\beta=0$) quantization in which the roles of the leading and subleading fall-offs are reversed \cite{Breitenlohner:1982jf,Breitenlohner:1982bm}. In either case, that this state of the ABJM theory should be characterized by a deformation of the theory by a source for the scalar operator is surprising. 

From an eleven dimensional perspective, this solution belongs to a class of spinning $M2$-branes very analogous to the continuous distributions of $D3$-branes discussed in \cite{Freedman:1999gk}.  As in that case, this class contains a privileged BPS configuration of $M2$-branes that is natural to associate holographically with a state on the Coulomb branch of the ABJM theory. From a four dimensional perspective, this  configuration (called the ``2+0--charge" solution in \cite{DeWolfe:2014ifa}) involves a bulk scalar that also appears in the 3--charge black brane solution. That the Coulomb branch solution spontaneously breaks the global symmetry of the boundary theory is reflected holographically in the four dimensional scalar containing only a normalizable fall off near the boundary. Since the 3--charge black brane can be reached from this Coulomb branch solution merely by tuning the rotation parameters of the $M2$ branes,  it seems likely that despite appearances, the bulk scalar profile of the 3--charge 
solution also encodes only a spontaneous symmetry breaking.

A possible resolution to this puzzle appears in \cite{Freedman:2013ryh,Freedman:2016yue}. There the authors demonstrate that by incorporating a manifestly SUSY covariant holographic renormalization scheme, a finite counterterm cubic in the scalars effectively shifts the identification of the source. Explicitly, in the conventions of the present work, they supplement the standard holographic counterterm action with a term of the form
\begin{equation}
S_{\mathrm{finite}}= \frac{1}{3\sqrt{3}}\int \dd^3x \,\sqrt{-\gamma}\,\phi^3.
\end{equation}
Analyzing the variations of the corresponding on-shell action leads to a quantization condition modified by the finite counterterm, which can be written
\begin{equation}\label{eq:impQNT}
\frac{\beta}{\sqrt{3}}-\frac{\alpha^2}{6} = 0.
\end{equation}
For further details, see appendix \ref{app:HRG}.

In \cite{Freedman:2013ryh}, the authors find that such an improved quantization condition enables a succesful match between calculations of the ABJM free energy on $S^3$ performed both holographically and from the field theory using localization techniques.  For our purposes, we note that when (\ref{eq:impQNT}) is applied to the asymptotic behavior of the bulk scalar (\ref{eq:bkgSFO}), the dual source for the scalar operator vanishes identically. This is in line with our $M2$-brane intuition developed above. Consequently, we will adopt these boundary conditions for the bulk scalar, and interpret the dual ABJM phase as a finite density state of strongly coupled matter in which a scalar operator has spontaneously acquired an expectation value.

We will primarily be interested in this phase at very low temperatures. In this case, the entropy density at fixed chemical potential (as well as the specific heat $C_\mu$) vanishes like
\begin{equation}
s \approx \frac{16}{3}\pi^2\mu\, T +\mathcal{O}(T^3).
\end{equation}
The fact that the entropy vanishes at zero temperature is reflected in the absence of an extremal horizon in this limit. Instead, at $T=0$ the geometry becomes singular in the IR. This singularity is of the ``good" kind in the classification of \cite{Gubser:2000nd}, and it is expected that the singularity is resolved during the oxidation to eleven dimensions.  A partial lift of this solution was discussed in \cite{DeWolfe:2014ifa}, where it is shown that the singular IR geometry can indeed be resolved to a (non-singular) $AdS_3\times \mathbb{R}^2$ region in one higher dimension. 

That the singularity is ``good" further implies that the zero temperature solution is continuously connected to a branch of charged black brane solutions. Thus, we anticipate that we can uncover features of the zero temperature background by taking a small black brane limit along this branch. In what follows, we will employ such a strategy for both computational convenience as well as to help clarify our results.

Another useful susceptibility controlled by the background solution is the uniform static charge susceptibility, $\chi_0$, defined at fixed temperature as
\begin{equation}
\chi_0\equiv \left( \frac{\partial\rho}{\partial\mu}\right)_T=\frac{2}{3}\frac{8\pi^2 T^2+9\mu^2}{\sqrt{16\pi^2 T^2+9\mu^2}}.
\end{equation}
At low temperatures, this susceptibility behaves like
\begin{equation}\label{eq:ssBKG}
\hat{\chi}_0 \approx 2+\frac{64}{81}\pi^4\hat{T}^4+\mathcal{O}(\hat{T}^6),
\end{equation}
where hatted quantities are dimensionless ratios of the un-hatted quantity and the chemical potential. The static charge susceptibility will play a central role in our investigation of the dual finite density phase of ABJM matter, and we will discuss this quantity in considerable detail below.

\section{Fluctuations and Linear Response}\label{htl}
The thermodynamic analysis of the preceding section reveals several interesting broad stroke features of the 3--charge black brane's holographic dual. In particular, the vanishing entropy density at zero temperature encourages questions about the nature of low energy excitations in the dual phase of matter. From the perspective of the field theory, a natural way to investigate the properties of these putative excitations is by way of the correlation functions for field theory operators. For example, in linear response theory considerable mileage can be gained from retarded Green's functions, which can be used to construct various spectral functions, conductivities, and susceptibilities. 

Holography provides an extremely efficient means of obtaining such retarded Green's functions. In what is by now a textbook application of gauge/gravity duality, the field theory calculation is reformulated in the gravitational language as a boundary value problem for linearized bulk fluctuations (while not a textbook, a useful exposition appears in \cite{Kaminski:2009dh}). In a top down holographic approach such as the one we follow here, one can identify both fermionic and bosonic fluctuations whose dynamics is entirely determined by the underlying gauged SUGRA. These fluctuations are then related holographically to two point functions for various fermionic and bosonic operators of the ABJM theory, respectively. 

Our immediate goal is to understand to what extent some interesting singularities in fermion spectral functions computed holographically in the 3--charge black brane can be related to features in the retarded density-density correlation function. The fermion spectral functions have previously appeared in the literature \cite{DeWolfe:2014ifa}. We will quickly summarize the relevant results, before turning our attention to the density-density correlator and the computation of the non-uniform static susceptibility, which are new.

\subsection{Fermion Spectral Functions}\label{sec:num_construction}
The 4$D$ $\mathcal{N}=8$ gauged SUGRA \cite{deWit:1982ig} contains 56 spin-1/2 fermions, and 8 gravitini. In any given bosonic background, fluctuations of these fermionic modes will generically couple to one another. This mixing greatly complicates the analysis of linearized fermionic fluctuations in supergravity theories. 

Fortunately, for particular backgrounds built from bosonic fields that are invariant under some subgroup $H\subset SO(8)$, the situation is often much improved. As we have seen above, the 3--charge black brane is a solution within a truncation of the SUGRA to bosonic fields invariant under the $H = U(1)^4$ Cartan of $SO(8)$. Thus, by studying the weight vectors of the SUGRA fermions, one can easily distinguish subsets of the spin-1/2 fermions which can not mix either amongst themselves, or with the gravitini. For further details on this disentangling, we refer the reader to \cite{DeWolfe:2014ifa}.

Solving for these decoupled bulk fermions allows one to holographically compute Green's functions of fermionic operators ``$\mathcal{O}_\psi$" in the ABJM theory. Because the dual phase of matter is at finite density, one can wonder whether or not the dual fermions which are charged under the active chemical potential(s) form Fermi surfaces. The existence of such a Fermi surface can be diagnosed from the analyticity properties of the fermion two point functions. On very general grounds \cite{Faulkner:2009wj,Cubrovic:2009ye,Gubser:2012yb}, at low energies (measured from the chemical potential) the holographic fermion Green's function in the vicinity of a bulk zero mode can be written \begin{equation}
G_R^{\mathcal{O}_\psi\mathcal{O}_\psi}(\omega,k) = \frac{Z}{\omega-v_F(k-k_F) -\Sigma(\omega)},
\end{equation}
which mimics the familiar Landau Fermi liquid parametrization. Unlike the Landau Fermi liquid, the quasi-particle weight $Z$ need not remain finite at the Fermi surface, nor does $\Sigma$ necessarily contain subleading contributions. In fact since $\Sigma$ is in general complex, it always provides the leading contribution to the width of any fermionic excitations. In any case a Fermi surface at momentum $k_F$ appears as a simple pole in the retarded fermionic Green's function at $\omega=0$, signalling the presence of gapless fermionic modes at $k=k_F$.

By constructing these Green's functions holographically, the authors of \cite{DeWolfe:2014ifa} indeed find Fermi surface singularities dual to bulk finite momentum fermion zero modes. Moreover, by studying the properties of the Green's function (and its related spectral function) at finite frequency, it was noted that in the 3--charge black brane background these holographic  Fermi surfaces are accompanied by perfectly stable fermionic excitations at low energies.

\begin{figure}
\centering
\includegraphics[scale=0.35]{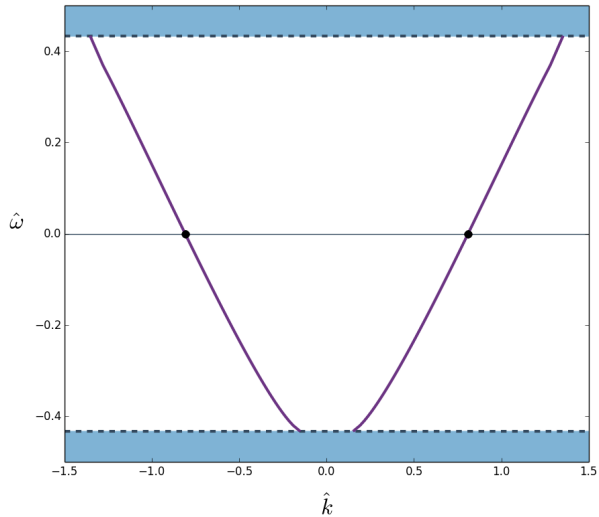}
\caption{\label{fig:disp} The spectral function $A(\omega,k)$ computed from ``Class 2" fermions in the 3--charge black brane  background. The solid lines indicated delta function peaks in the spectral weight. This weight broadens in the shaded bands, and has been omitted from the plot for clarity.}
\end{figure}

To characterize these excitations, one can compute the spectral function from the imaginary part of the correlator:
\begin{equation}
A(k,\omega) = \frac{i}{2}\Tr \,({\bf G}_R^{\psi\psi} - {\bf G}_R^{\psi\psi}\,^\dagger).\end{equation}
The trace is taken over the spinor indices of the retarded Green's function, which renders the spectral function rotationally invariant. In figure \ref{fig:disp} an example of this spectral function computed holographically from the 3--charge black brane is shown. This particular spectral function corresponds to the ABJM fermions which display the largest fermi surface of those studied in this state in \cite{DeWolfe:2014ifa}. The spectral function is characterized by a region of infinitely long-lived fermionic excitations with a dispersion relation marked by the solid purple line. The Fermi surface is marked by large black dots, and is at $\hat{k}_F\equiv k_F/\mu\approx 0.81$. 

For fluctuations with $|\omega/\mu|>\sqrt{3}/4$ (shaded bands in the figure) the normal mode pole in the correlator moves off the real axis and the fluctuations acquire a finite width. The fact that this kinematic regime of stability is present in these spectral functions led the authors of \cite{DeWolfe:2014ifa, DeWolfe:2013uba} to associate the stable region with a gap in which modes mediating interactions with the fermionic excitations are absent.  

The holographic picture that emerges is thus that of a finite density state in which the strongly coupled fermionic  degrees of freedom dual to ``Class 2" SUGRA modes appear to have organized into a Fermi surface. The low energy fermionic excitations around this Fermi surface behave similarly to those in a Fermi gas: perfectly stable modes with linear dispersion near the Fermi surface.  

\subsection{The Holographic Static Susceptibility Setup}\label{sec:statsus}
As discussed in the previous subsection, the holographic fermion spectral function singularities identified in the SUGRA fluctuation analysis can be interpreted as signaling the existence of a Fermi surface built from ABJM fermions carrying charge under the global $U(1)$ number densities. In an attempt to better understand these singularities, it is natural to wonder whether or not they leave their mark in other field theory observables.

An obvious contender for one such observable is the density-density correlator $G_R^{\rho\rho}(\omega,k)$. In the static limit ($\omega\to 0$), this correlation function quantifies the  static susceptibility of the state. In other words, as a function of momentum the static susceptibility measures the response of the charge density to a small amplitude, stationary charged perturbation. 

To compute the static susceptibility holographically, we  study fluctuations of the time component of the bulk gauge field subject to the appropriate boundary conditions. In the classical background of the 3--charge black brane, linearized fermionic and bosonic fluctuations clearly decouple. However, the symmetries of the background do not permit one to fully decouple the set of bosonic fluctuations, and these modes will generically mix amongst each other. 

As the bulk scalar is uncharged under the gauged $U(1)$, on general grounds one anticipates a total of 2+2+1 = 5 bosonic gauge invariant modes. The obvious $SO(2)$ isometry of the background will be broken by any finite momentum fluctuation. If we take this momentum to be along the $x$-direction, then the gauge invariant fluctuations can be organized into representations of the unbroken\footnote{Note that although the gauged SUGRA action contains a topological term proportional to $F\wedge F$, this term couples to the axions. These axions vanish in the 3--charge black brane solution, and thus this term does not affect the linearized fluctuation analysis.} $\mathbb{Z}_2\subset SO(2)$ which labels the parity of the mode under $y\to-y$. Fluctuations of the temporal component of the gauge field are clearly even under this $\mathbb{Z}_2$, and will mix with the two other parity even modes. We collectively denote these modes $\mathcal{Z}_+$. 

To construct explicitly the gauge invariant  $\mathcal{Z}_+$, it is convenient to first write the bulk fields as background plus fluctuation
\begin{equation}\label{eq:bkgdPLUSfluct}
g = \bar{g} + h \qquad A = \bar{A}+a \qquad \mathrm{and}\qquad \phi = \bar{\phi}+\varphi
\end{equation}
and to adopt an appropriate plane wave ansatz for both the fluctuations $\psi_I = \{h,a,\varphi\}$ and the gauge transformation parameters $\epsilon_I = \{\xi,\lambda\}$. Here $\xi$ represents a diffeomorphism and $\lambda$ denotes a $U(1)$ gauge transformation. Schematically, 
\begin{equation}\label{eq:planeWave}
\psi_I (t,r,x,y)\sim \psi_I(r)e^{i(kx-\omega t)}\qquad \mathrm{and} \qquad \epsilon_I\sim \epsilon_I(r)e^{i(kx-\omega t)}.
\end{equation}

By studying the action of the Lie derivative along $\xi$ and the $U(1)$ gauge transformations parametrized by $\lambda$ on these plane wave fluctuations, it is easy to construct modes that are invariant under both. In the static limit, one choice is given by $\mathcal{Z}_+ = \{\mathcal{Z}_s,\mathcal{Z}_1,\mathcal{Z}_2\}$ where
\begin{align}
\mathcal{Z}_s &=\varphi - \frac{\bar{\phi}'}{\bar{g}_{yy}\,'}h_{yy}\label{eq:sZs}\\
\mathcal{Z}_1 &= \,h_{tt}-\frac{\bar{g}_{tt}\,'}{\bar{g}_{yy}\,'} h_{yy}\label{eq:sZ1}\\
\mathcal{Z}_2 &=\,a_t - \frac{\Phi'}{\bar{g}_{yy}\,'}h_{yy}\label{eq:sZ2}
\end{align}
and a $'$ has been used to denote a radial derivative.

In our analysis, the gauge invariant $\mathcal{Z}_+$ are primarily important  because they demonstrate how to properly impose consistent boundary conditions on the bulk fluctuations. In practice, we work directly in terms of the $\psi_I$, changing variables slightly so that $h=e^{2\chi}\tilde{h}$. A technical point, examined in more detail in the appendix, is that the radial coordinate in which the background solution (\ref{eq:bkg})  is presented does not coincide with the Fefferman-Graham coordinate $\sr$ near the boundary. This slightly complicates the identification of ``normalizable" and ``non-normalizable" modes of the various fluctuations.  

The upshot of this boundary analysis is that for the bulk fluctuations with near boundary behavior
\begin{align}\label{eq:FOs}
\varphi =& \frac{f_1}{r}+\frac{f_2}{r^2}+\ldots\nonumber\\
\tilde{h}_{tt} = & ht_0 + \ldots\nonumber\\
\tilde{h}_{yy} = & hy_0 + \ldots\nonumber\\
a_t = & a_0 + \frac{a_1}{r}+ \ldots
\end{align}
a solution which corresponds holographically to turning on a source for {\it only} the charge density must obey the UV ($r\to \infty$) boundary conditions
\begin{equation}\label{eq:UVbcs}
f_2 +\frac{Q}{4}f_1 = (ht_0+hy_0) = 0\qquad \mathrm{and} \qquad  a_0 = 1.
\end{equation}

In order to better understand the analytic structure of the Green's function we compute holographically, it is advantageous to construct this correlation function at finite temperature, and study its behavior in the limit $\hat{T}\to 0$. From the gravitational perspective, this implies that we will be interested primarily in perturbations to background solutions with a horizon. In this case, the correct IR ($r\to r_H$) boundary condition is regularity of the fluctuation at the horizon. Again, further details are provided in an appendix.

Given a solution to the fluctuation equations of motion that is both regular at the horizon and satisfies (\ref{eq:UVbcs}) at the boundary, a straightforward application of the holographic dictionary gives the static susceptibility as
\begin{equation}
\chi(k)=-\lim_{\sr\to\infty}\sqrt{-g}\,\frac{\partial_\sr \,a_t}{a_t} = \left(a_1-\frac{1}{2}(Q+r_H)\sqrt{Q(Q+r_H)}\,hy_0 \right).\label{eq:ssus}
\end{equation}
We now turn to the computation of this susceptibility, and a subsequent discussion of our results.

\section{Results}\label{sec:rez}
\begin{figure}
\centering
\includegraphics[scale=0.52]{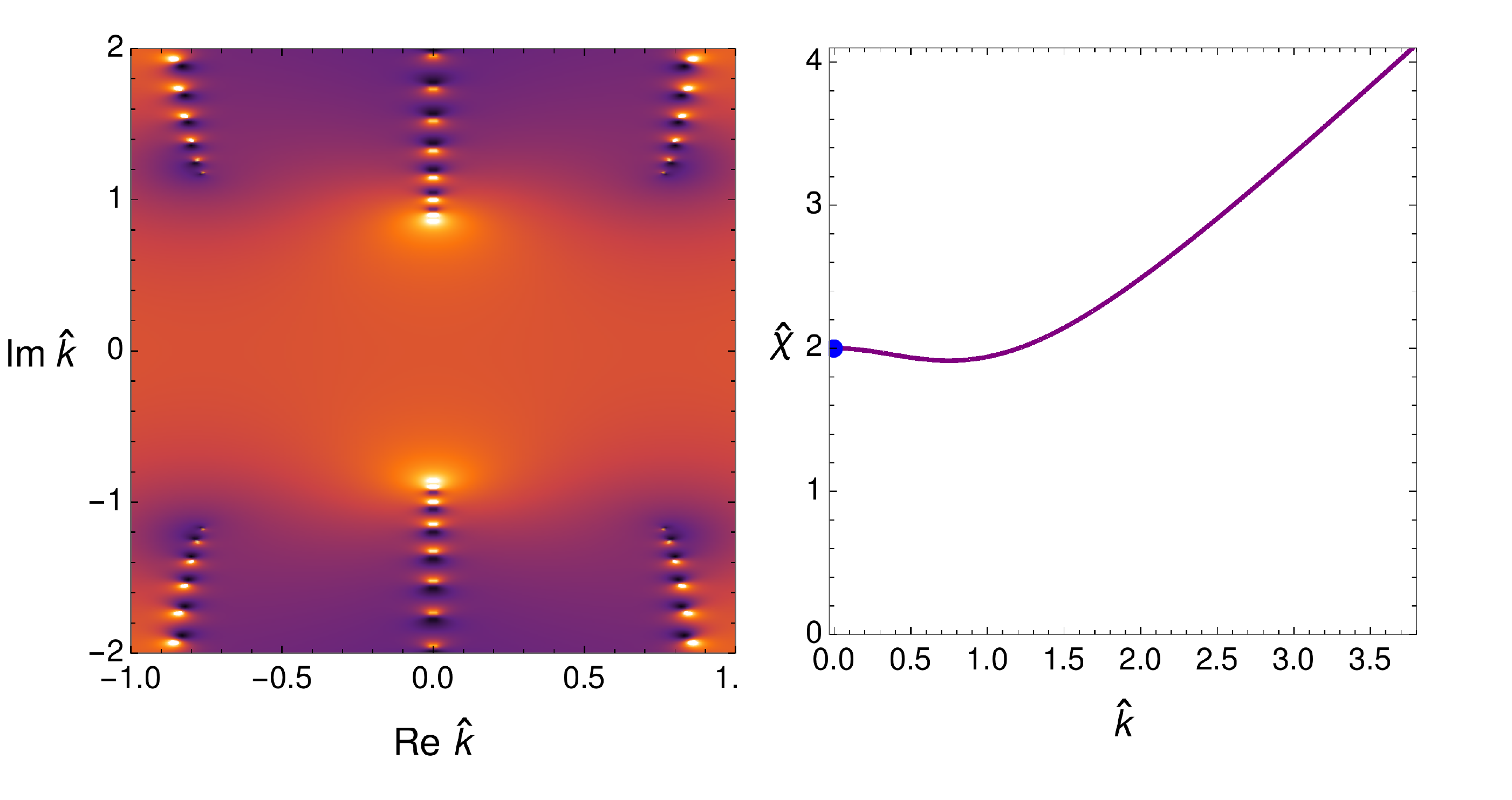}
\caption{\label{fig:ssRek} The low temperature static susceptibility along the real $\hat{k}\equiv k/\mu$ axis (right) and in the complex $\hat{k}$--plane (left). The susceptibilities pictured are computed at a temperature $\hat{T} =  10^{-5}$. In the rightmost plot, a point at $\hat{k}=0$ shows the location of the uniform static susceptibility computed from the background, in excellent agreement with the fluctuation analysis. }
\end{figure}

Our main results are summarized in figure \ref{fig:ssRek}. The rightmost plot shows the static susceptibility evaluated at real momenta. Importantly, the zero momentum limit is shown to be in excellent agreement with (\ref{eq:ssBKG}), demonstrating consistency between the fluctuation analysis and the background thermodynamics. For momenta large compared to the chemical potential, the fluctuations are primarily supported at large radii (near the boundary) and thus inherit the conformal characteristics of the UV fixed point. This is manifest in the linear rise of the susceptibility at large $\hat{k}$.

The leftmost plot of the same figure shows clearly the non-analyticities present in the static susceptibility at low temperature. At finite temperature these non-analyticities appear as a discrete set of poles which coalesce as the temperature is lowered. At zero temperature, they trace branch cuts in the complex momentum plane, which terminate at branch points naturally distinguished by whether or not $\mathrm{Re}\, \hat{k} = 0$.

\begin{figure}
\centering
\includegraphics[scale=0.36]{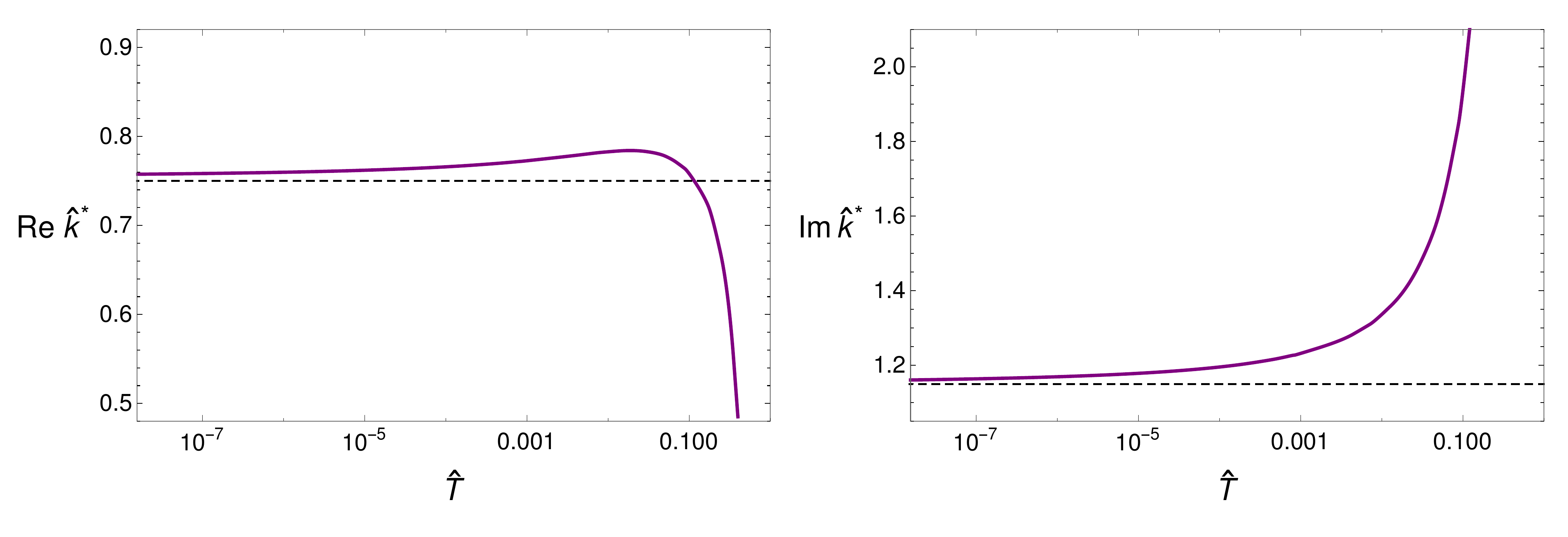}
\caption{ The location of the branch points with $\mathrm{Re}\, \hat{k} \ne 0$ as a function of temperature. The dashed lines are given by branch points of the scaling exponent $\nu_{||}^-$, to be discussed in more detail below.\label{fig:bpMtn}}
\end{figure}

By eye, the branch points with $\mathrm{Re}\, \hat{k} \ne 0$ appear to lie very close to the location $\hat{k}_F \approx 0.81$ of the Fermi surface identified via the fermion spectral functions evaluated in the same state. This is again reminiscent of the results from \cite{Blake:2014lva}, in which branch points in the susceptibility occur at ${\mathrm{Re}}\,\hat{k}^\star \approx \hat{k}_F$. As in that case, however, the match is not exact. In figure \ref{fig:bpMtn}, the location of the branch points corresponding to the $\mathrm{Re}\, \hat{k} \ne 0$ class is plotted against temperature. Evidently, at low temperatures these branch points approach $\hat{k}^\star = |\hat{k}_R|+i|\hat{k}_I|\approx 0.75 + 1.15\,i$. As the temperature is increased beyond $\hat{T}\sim 0.1$, the real part of the branch point quickly decreases. This behavior is anticipated from the analogous calculations in $AdS$--RN, in which it was shown that for temperatures large compared to the chemical potential the singularities in the 
static 
susceptibility are aligned along the imaginary $\hat{k}$ axis.

Perhaps most interestingly, figure \ref{fig:ssRek} indicates a second class of branch cut persisting at very {\it low} temperatures along the imaginary $\hat{k}$ axis. This feature was absent in the state dual to $AdS$--RN, and has important implications. Most notably, the magnitude of the imaginary part of these branch points is smaller than those at finite ${\mathrm{Re}}\, \hat{k}$. This observation, coupled with the fact that these branch points have $\mathrm{Re}\, \hat{k}^\star = 0$ signals an absence in long distance oscillations in the charge density induced by a charged perturbation. 

\begin{figure}
\centering
\includegraphics[scale=0.43]{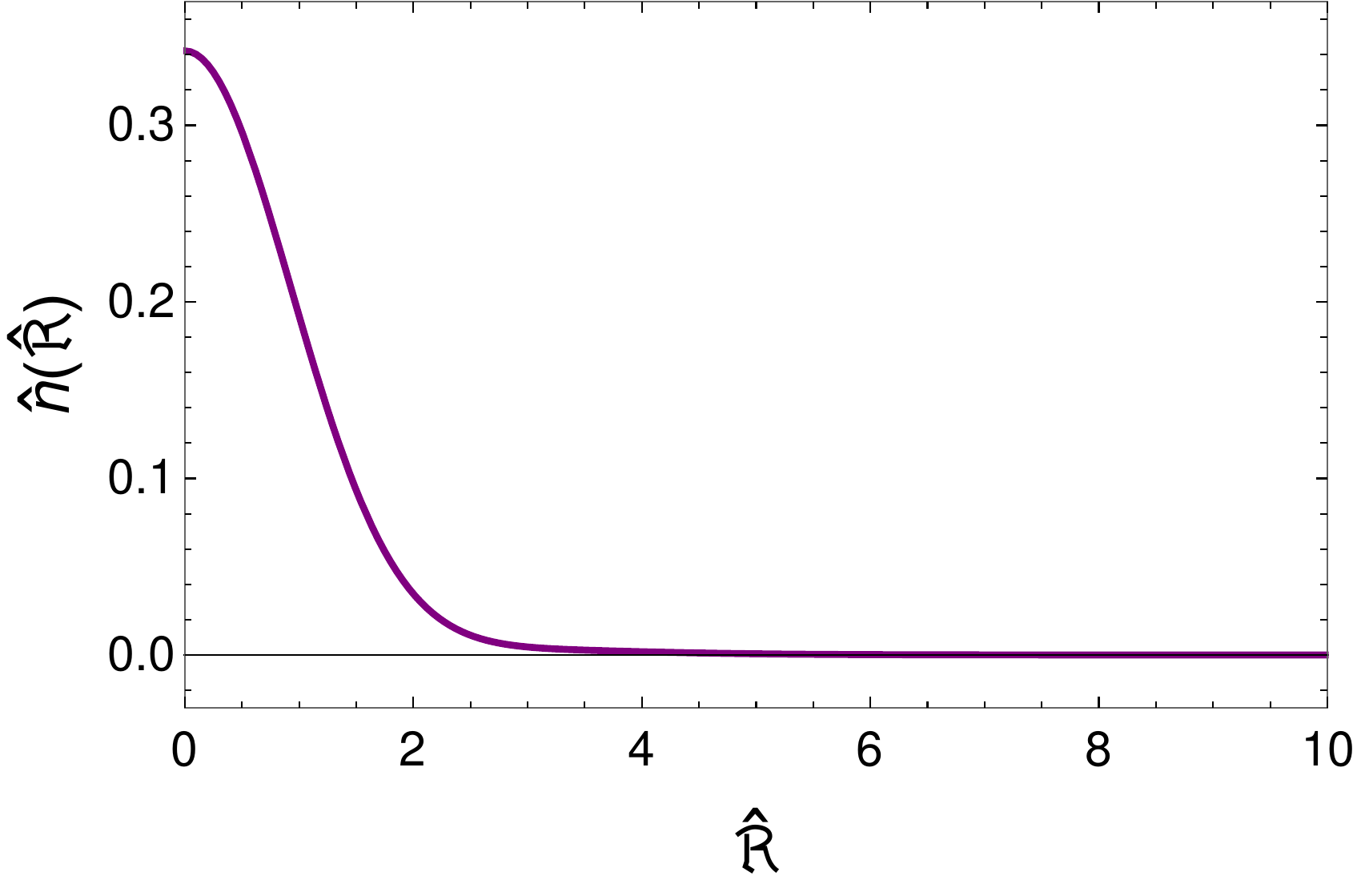}
\includegraphics[scale=0.43]{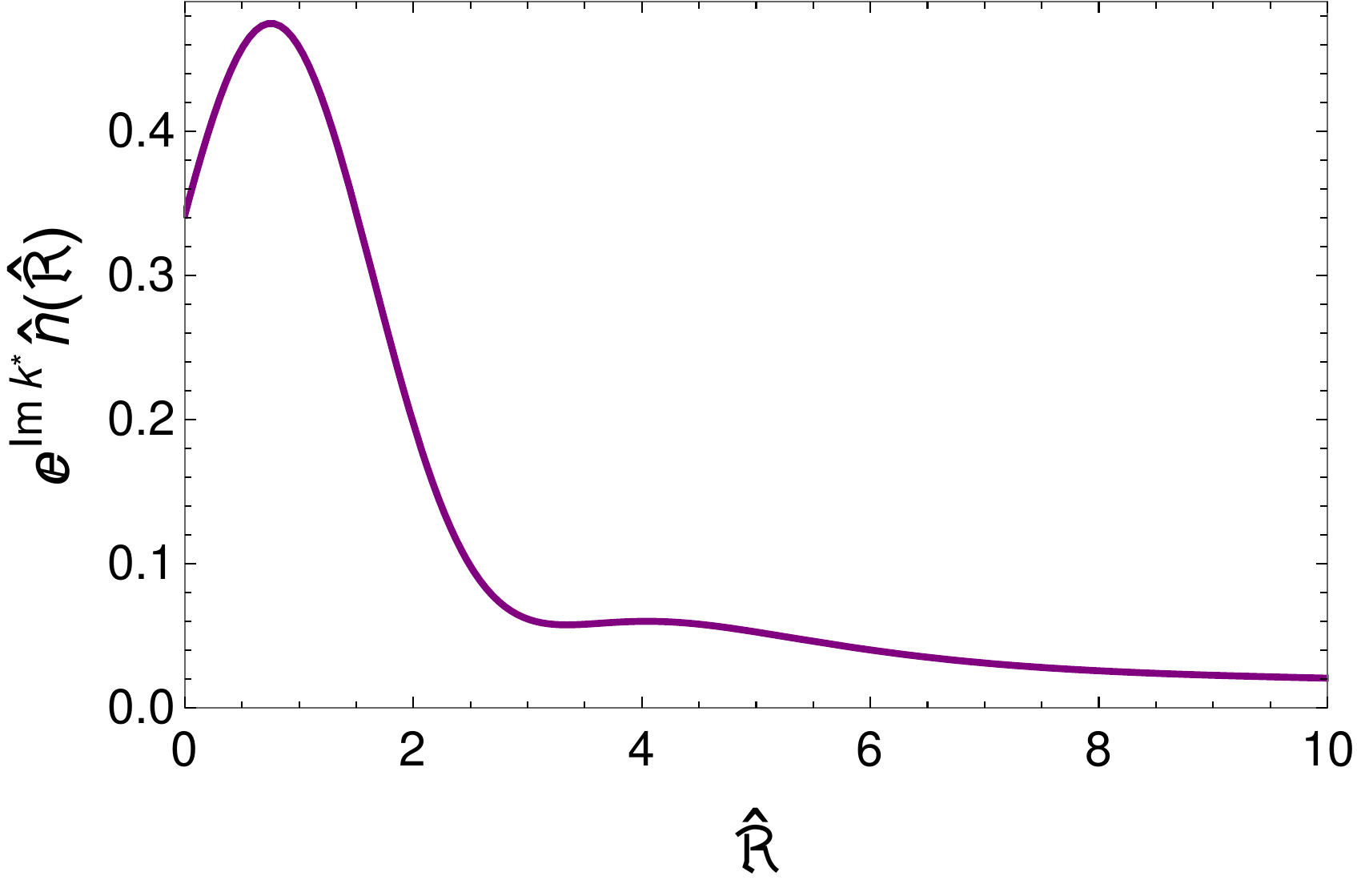}
\caption{ The linearized charge density response due to the introduction of a charged impurity at low temperature , $\hat{T} = 10^{-5}$ (left). No oscillations are easily discernible at large distances. To observe the effects of the branch points with  $\mathrm{Re}\, \hat{k}^\star \ne 0$, it is helpful to remove the exponential damping from the branch point along the imaginary axis (right).\label{fig:indN}} 
\end{figure}

One can demonstrate this explicitly by studying this induced charge density $n$ in the linear response limit of the gauge theory. In this limit, a charged impurity $\delta Q$ deforms the charge density like
\begin{equation}\label{eq:nFrie}
n(\mathfrak{R}) = \int \frac{\mathrm{d}^2k}{(2\pi)^2}e^{i k \mathfrak{R}}\,\chi(k)\,\delta Q(k).
\end{equation}
As long as the impurity is relatively well behaved, the induced charge density will be dominated by contributions to the integral coming from the branch points in the susceptibility. Choosing for the sake of simplicity a gaussian $\delta Q(\mathfrak{R}) = e^{-\mathfrak{R}^2/2}$,  the low temperature induced charge density is readily obtained by numerically integrating (\ref{eq:nFrie}), and the result is plotted in figure \ref{fig:indN}. Here $\mathfrak{R}$ is a radial coordinate in the boundary theory, and should not be confused with the bulk radial direction. Unlike the state dual to $AdS$--RN, at large distances ($\mathfrak{R}\gg 1)$ the response is dominated by the branch point on the imaginary axis and no analogue of Friedel oscillations are present.

\section{Discussion}\label{disc}
\subsection{Singularities and Scaling Exponents}
Given the slight mismatch between the poles of fermion two point functions and the non-analyticities present in the static susceptibility, it would be advantageous to have an analytical handle on the heritage of one (or both) of these features. A striking observation made by \cite{Blake:2014lva} was an apparent connection between the branch points in their numerical result and those of the ``semi-local quantum critical scaling exponent'', $\nu^-_k$.

\begin{figure}
\centering
\includegraphics[scale=0.38]{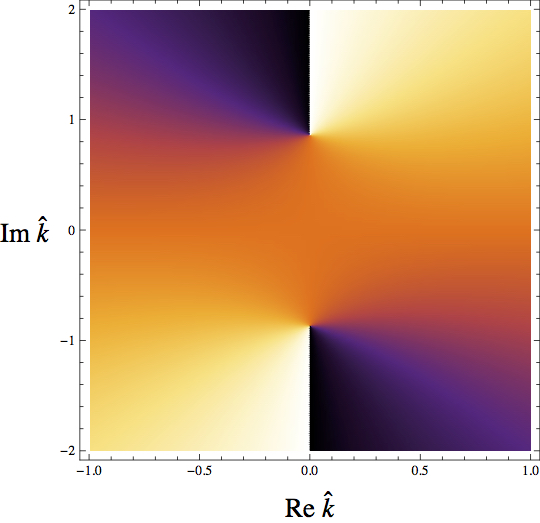}
\includegraphics[scale=0.38]{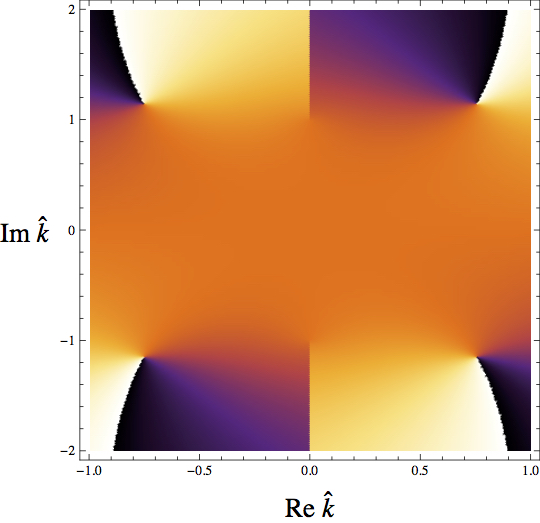}
\caption{ The analytic structure of $\nu_{||}^0$ (left) and $\nu_{||}^-$ (right) in the complex momentum plane. The plots show the argument of the scaling exponent, which clearly reveals the existence of two classes of branch points. Both of these classes are present in the susceptibility. Numerically, they occur at $\hat{k}^\star \approx \pm 0.87\, i$ and $\hat{k}^\star \approx \pm 0.75 \pm 1.15\, i$ in excellent agreement with the numerically computed susceptibility.\label{fig:nuArgs}}
\end{figure}

Although the 3Q-black brane lacks an extremal horizon, it nevertheless inherits some of the interesting properties of extremal $AdS$--RN as a consequence of an IR geometry that is conformal to $AdS_2\times \mathbb{R}^2$. Importantly, \cite{Anantua:2012nj} demonstrated that longitudinal fluctuations of the gauge field in this IR geometry are characterized by a set of three scaling exponents that generalize the $\nu^-_k$. In the conventions we adopt here, these exponents are given by
\begin{align}
\nu_{||}^0(k) = &\, \sqrt{1+\frac{4}{3}\hat{k}^2}\nonumber\\
\nu_{||}^\pm(k)= &\, \frac{1}{\sqrt{3}}\sqrt{11+4\hat{k}^2\pm 8\sqrt{1+\hat{k}^2}}. \label{eq:3Qnuks}
\end{align}
The exponents $\nu_{||}^0$ and $\nu_{||}^-$ have particularly interesting branch points at 
\begin{equation}
\hat{k}^\star=\pm i\frac{\sqrt{3}}{2}\qquad \mathrm{and}\qquad \hat{k}^\star = \left(\frac{57}{16}\right)^{\frac{1}{4}}e^{\frac{i}{2}(\pm \pi \pm \mathrm{arctan} \frac{4}{\sqrt{3}})}
\end{equation}
 respectively. The $\pm$ signs in the branch points of $\nu_{||}^-$ are not correlated (i.e. there are four such branch points). The argument of these scaling exponents is plotted in figure \ref{fig:nuArgs}, where one can note very good agreement between the numerically obtained analytic structure of static susceptibility at low temperatures and the analytically derived scaling exponents in (\ref{eq:3Qnuks}).

The similarity between $\mathrm{Re}\,\hat{k}^\star$ from $\nu_{||}^-$ and the large $k_F$ from the fermion spectral functions in the $AdS$--RN and 3Q black brane backgrounds is surprising on several levels. Foremost, from the perspective of the linearized fermion zero mode calculations, the existence and location of a holographic Fermi surface depends intimately on the details of the couplings in the bulk Dirac equation. These couplings in turn are generically functions of the background fields. Changing these couplings by $\mathcal{O}(1)$ deformations can radically alter the analytic structure of the fermion spectral function. However the holographic computation of the static susceptibility is agnostic to the details of these couplings, as it obviously depends only on a bosonic subsector of the linearized fluctuations. That these two computations should yield results that possess any commonalities is therefore rather unexpected from the supergravity perspective. We return to this point shortly.

Moreover, the $\nu_{||}$ that characterize the static susceptibility are not the same as the analogous exponents that govern the IR properties of the fermion correlation functions. Nor do they share the same analytic structure in the complex momentum plane. In the extremal $AdS$--RN background, for example, spin-1/2 supergravity spinors that allow for zero modes are controlled in the IR by a scaling exponent of the form
\begin{equation}
\nu_F = \sqrt{2\hat{k}^2-\frac{1}{12}}.
\end{equation}
This exponent clearly has branch points along the real $\hat{k}$ axis, and in fact hints at the existence of an instability of the extremal solution to the formation of an ``electron star'' (see e.g. \cite{Hartnoll:2011fn}). More relevant to the current investigation,  these branch points occur at $\hat{k}^\star = \pm1/(2\sqrt{6})$, which does not seem to have much to do with the Fermi momentum---especially when compared to the branch points of $\nu^-_k$ evaluated in the same state (see discussion around (\ref{eq:nu_kAdS})).

Perhaps even more dramatically, analogous supergravity spinors in the 3Q--black brane do not readily appear to possess any such scaling exponent in the deep IR. Their behavior is instead dictated by the appearance of a momentum independent ``gap" \cite{DeWolfe:2014ifa} in the static limit. In neither the 3Q--black brane nor the extremal $AdS$--RN solution are fermion Dirac equations influenced by the $\nu_{||}$ in their IR limit. 

Thus we arrive at one of the most important lessons of our calculation: at leading order in the strong coupling, large $N$ limit, certain fermionic {\it and} bosonic correlation functions in these finite density states of the ABJM theory  appear to identify very similar length scales $1/k_F \approx 1/\mathrm{Re}\, k^\star$ in holographic computations that are outwardly independent of one another. The details of these correlation functions---a fermion spectral function in one case, and a static susceptibility in the other---encourage an identification between this length scale and the existence of a Fermi surface in the ABJM phase, lending some support to the picture advocated in \cite{DeWolfe:2011aa}. We will criticize this identification shortly.

The present calculation illuminates several other interesting features of holographic matter. Among these, it is somewhat curious that the addition of the bulk scalar to the gravitational theory is at once responsible for opening of a new channel of fluctuation (that associated to $\nu_{||}^0$) which destroys the long distance oscillations in the charge density, while simultaneously allowing for a stable region in the fermion spectral function where infinitely long-lived fermionic excitations propagate.  Naively, one might expect that such a stable region would enhance charge oscillations at long distance, but we find this is not the case. 

It is important to emphasize that the majority of the lessons we have learned were a consequence of our adherence to a ``top-down" implementation of the holographic dictionary. By working with bulk equations of motion that are fully constrained by the underlying supergravity theory, we can confidently work within a consistent holographic framework. In such a setup, details of the bosonic background are communicated to the fermionic fluctuations via highly tuned (background dependent) couplings in the bulk Dirac equations. It is only in this context that we can attempt quantitative matches between the features of fermion spectral functions and static susceptibility in states of ABJM theory. 

A trivial manifestation of this was the importance of identifying the correct boundary conditions for the fluctuations of the bulk scalar. This identification was especially noteworthy in the present setting, as the quantization of the scalar was influenced by the existence of finite boundary counterterms imposed by the supersymmetry of the boundary theory. Altering these boundary conditions artificially would have led to a mismatch both in the uniform ($\hat{k}\to0$) limit of the static susceptibility, as well as the location of the branch points identified numerically relative to those of the $\nu_{||}$.

\subsection{Commentary}

If any connection between the branch points of the $\nu_{||}$ and the Fermi surface poles in the fermionic two-point functions is to hold, then there is some tension to be resolved. On one hand, it is no great surprise that the singularities in both correlation functions are $\mathcal{O}(\mu)$ in all examples. At zero temperature, the chemical potential provides the only relevant scale in the background. This is then roughly the scale at which  bulk fluctuations distinguish between IR and UV portions of the geometry. For static fluctuations with momenta sufficiently large compared to the chemical potential, the fluctuation probes the UV $AdS_4$, a conformally invariant critical point. In the vicinity of this point, static correlation functions are analytic at non-zero momentum. Accordingly, if any zero modes are to arise one typically expects them 
to do so at small to  moderate values of momentum in units of the chemical potential. Thus, while it seems unlikely to us, we can not currently rule out the possibility that the ``$1k_F$" singularities occur at approximately the same  momentum simply by numerical coincidence. An interpretation in terms of such a coincidence would be in line with the expectations of the ``cohesive charge density" picture of the holographic Fermi surface briefly summarized in the overview.

If we instead entertain the possibility that the singularities in the fermion correlators and the charge susceptibility {\it are} related, then why might it be that  ${\mathrm{Re}}\, \hat{k}\approx k_F$ as opposed to ${\mathrm{Re}}\, \hat{k}= k_F$? One possibility is that this effect can be attributed to operator mixing between the charge density and the energy and pressure densities,  along with (in the present case) the spinless operator dual to the bulk scalar. This mixing is captured holographically by couplings between the linearized fluctuations of the corresponding bulk modes.  Such couplings can in principle open new modes as well as deform normal modes that were present in the decoupled system, shifting their eigenvalues. Some examples of this well known phenomenon with a holographic lean appear in e.g. \cite{Kaminski:2009dh, DeWolfe:2015kma}.

On the other hand, as reviewed in the introduction, the ``2" in $2k_F$ is essentially a geometric consequence of gapless excitations at a sharp Fermi surface. What might it mean for holographic response that the branch points in the static susceptibility have a real part that is closer to $k_F$ than $2k_F$ from the perspective of fermionic correlators in the same state?

 One possibility is that the ``Fermi surface singularities" in the fermion two-point functions are not in fact indicative of a sharp Fermi surface in the dual phase.  This could be the case if, for example, the zero mode they detect is actually due to an ABJM scalar that is bound to an ABJM fermion in the gauge invariant operator dual to the bulk fermion. In the schematic classification of \cite{DeWolfe:2014ifa}, for example, the ``Class 2" bulk fermions emphasized in this work are dual to a gauge theory operator of the form $\mathrm{Tr}\lambda_1 Z_{i}$ for $i=2,3,4$. In this labeling, $\lambda$ is a complex combination of gauge theory fermions while the $Z$ are complex scalars. 

In charged black brane backgrounds, scalar zero modes are very often marginal modes that sit precisely at the boundary of a spatially modulated instability. Examples of this behavior are plentiful---some noteworthy realizations include \cite{Gubser:2000ec, Donos:2011bh, Faulkner:2009wj}. However the bulk fermion zero modes do not typically suffer the same fate. As demonstrated in \cite{Faulkner:2009wj}, in contrast to the scalar instabilities, tuning $k$ away from $k_F$ does not in general result in the fermionic zero mode migrating into the upper half frequency plane. Thus, if the singularities in the fermionic correlators are actually reflecting the behavior of the $Z_i$, the dual ABJM phase would be characterized by gapless scalar modes gathered at the ``$1k_F$" momentum. This seems like a strange state of affairs in a translationally invariant phase that is stable against these modulated scalar fluctuations (at least in the large $N$ limit).

Another possibility is that while the $1k_F$ singularities in the fermion two-point functions {\it are} in fact signaling the existence of a sharp Fermi surface, the large bath of charged scalars radically alters the dominant interaction channels available to charged excitations. If the interactions were such that parallel patches of the Fermi surface were effectively inaccessible to one another, it is plausible that the susceptibilities would be dominated instead by wave vectors at or very near $k_F$. Alternatively, it might be that the important physics of the fermionic correlation function computed holographically is actually that of a ``two particle" correlator. Such an interpretation would appeal to the holographic identification of the bulk spin-1/2 mode with a boundary operator of the form $\mathcal{O}_\psi \sim \mathrm{Tr}\lambda Z$. In this light, bulk fermion zero modes might themselves simply be signals of $2k_F$ singularities in the ``$\mathcal{O}_\psi$ static susceptibility".  It would be very 
interesting to further explore the tenability of such phenomena in a toy model of interacting bosons and fermions outside of holography.

 In \cite{DeWolfe:2014ifa}, in addition to the large Fermi surface that held our attention above, a further two Fermi momenta were detected in the set of fermion spectral functions from the 3Q black brane. Evidently, no accompanying branch points arise in either $\nu_{||}^-$ or the static susceptibility. 

A sketch towards one possible resolution of this apparent discrepancy might begin with the following observation: the additional Fermi momenta identified in \cite{DeWolfe:2014ifa} are much smaller than the large Fermi surface pictured in section \ref{sec:num_construction}. Thus, by a Luttinger-like argument, it is natural to anticipate that these smaller Fermi surfaces, $k_F^<$, control only
\begin{equation}
\frac{g_f^<}{g_f^>}\cdot\frac{q^<}{q^>}\cdot\left(\frac{k_F^<}{k_F^>}\right)^2 = 3\cdot\frac{1}{3}\cdot\left(0.34\right)^2\approx 0.1
\end{equation}
of the fermionic charge density relative to the larger $k_F^>$. The multiplicative factor  of 3 accounts for the relative degeneracy of the Fermi surfaces, while the factor of 1/3 accounts for the differing charges of the ABJM fermions under the active $U(1)$. These factors are chosen to correspond to the larger of the two small Fermi surfaces. It is thus conceivable that the influence of the charged degrees of freedom around these smaller Fermi surfaces on the static susceptibility is completely overshadowed by the effects of the larger Fermi surface and/or other charged modes in the dual state.

Yet another curious observation is that in both the $AdS$--RN and 3Q--black brane backgrounds the non-analyticities in the static susceptibility appear to be controlled exclusively by the IR region of the geometry. On general grounds, this is not guaranteed to be the case. As emphasized in \cite{Anantua:2012nj}, the static susceptibility is holographically realized as a static fluctuation calculation, and is thus in principle sensitive to the entirety of the bulk geometry. This a priori leaves open the possibility for the existence of (normalizable) zero modes in the longitudinal fluctuations at real values of spatial momentum. Evidently such zero modes do not exist in these backgrounds. The presence of such zero modes would generically imply both long distance oscillations and power law decay in the induced charge density, and thus it would be very interesting to find holographic examples of this behavior.

To motivate this novel feature, it is perhaps instructive to examine the holographic susceptibility in a simpler setting in which it is not necessary to construct the fluctuations numerically. Within the $U(1)^4$ SUGRA truncation we are working in, there exist also a variety of uncharged zero temperature solutions.  While the phase dual to these ``RG flow" like backgrounds is not at finite density, the increased symmetries greatly simplify the fluctuation analysis and often allow one to compute various holographic correlation functions analytically. 

For concreteness, we turn our attention to the  horizonless ``2+0" charge background, given by
\begin{equation}
\dd s^2 =  U\left(-\dd t^2 +\dd \vec{x}^2\right)+\frac{\dd r^2}{U}\qquad \bar{\phi}(r) = \ln\left( 1+\frac{Q_1}{r}\right)\qquad \Phi_0 = \Phi_1 = 0
\end{equation}
with
\begin{equation}
U(r) = r^2 \left(1+\frac{Q_1}{r} \right).
\end{equation}
Turning on independent fluctuations $a^i$ of the two gauge fields kept in this ansatz, it is straightforward to repeat the steps  in section \ref{sec:statsus} and arrive at two simple and decoupled equations for the longitudinal fluctuations of the gauge fields:
\begin{align}
0 = & \,a^0_L\,'' +\frac{2}{r}a^0_L\,'-\frac{k^2}{r^2(Q_1+r)^2}a^0_L\\
0 = &\, a^1_L\,''+\frac{2}{Q_1+r}a^1_L\,'-\frac{k^2}{r^2(Q_1+r)^2}a^1_L
\end{align}
Note that because the background preserves Lorentz invariance in the field theory directions, the fluctuation equations (and their solutions) can be trivially generalized to include finite frequency as well. To make our point, however, we remain in the static limit. The solution to these equations most regular as $r\to 0$ is given by
\begin{equation}
a^i_L= \left(\frac{r}{Q_1+r}\right)^{-\frac{(-1)^i}{2}+\nu_k}
\end{equation}
with 
\begin{equation}
\nu_k = \sqrt{\hat{k}^2+\frac{1}{4}}
\end{equation}
and $\hat{k}\equiv k/Q_1$. Accordingly, the static susceptibility in the holographic phase dual to this background is
\begin{equation}\label{eq:2p0sus}
\chi(k) \propto Q_1\left(-\frac{(-1)^i}{2}+\nu_k\right).
\end{equation}

Although the dual state of the ABJM theory is at zero density, equation (\ref{eq:2p0sus}) realizes several of the features of the more complicated 3Q--black brane susceptibility computed numerically and presented in section \ref{sec:rez}. Most notably, there is a non-vanishing susceptibility for the addition of ``1-type" charge at zero momentum, and the only non-analytic behavior in the susceptibility is dictated by the branch points of the scaling exponent $\nu_k$. Unlike in the 3Q background, in this case there are no branch points with $\mathrm{Re}\,\hat{ k}^\star \ne 0$.

\subsection{Extensions}
It is also  interesting to wonder to what extent the somewhat rough connection we observed between fermion response and static susceptibility is present in other holographic phases of matter. One modest point in favor of such a connection appears in the (top-down) computation of fermion spectral functions in certain ground states of ABJM that break the global $U(1)$ associated with conserved charge density \cite{DeWolfe:2016rxk}. In stark contrast to analogous bottom-up calculations, it was found that such fermion spectral functions showed no sign of a Fermi surface in any of the states studied.  

Although the symmetry breaking was spontaneous in some instances and explicit in others, a common feature of the bulk description of the holographic phase was a horizonless geometry that took the form of a domain wall interpolating between the maximally symmetric $AdS_4$ in the UV and a distinct $AdS_4$ in the IR. Importantly, an IR $AdS_4$ does not permit fluctuations controlled by a momentum dependent critical scaling exponent---there is no analogue of $\nu_{||}$ in such geometries---and one might be tempted to guess that the static susceptibility computed holographically in such a background would show no non-analytic behavior at momenta with $\mathrm{Re}\,\hat{k}^\star \ne 0$. If such a scenario were born out by explicit computation it would provide further support for a link between poles in the fermionic two-point functions and certain branch points in the static susceptibility. 

Other avenues to pursue might include generalizing the analysis to move along the one parameter branch of solutions that interpolate between $AdS$--RN and the 3Q--black brane. These are the 3+1 Q--black brane solutions of \cite{Cvetic:1999xp} in which all four generators of the Cartan of $SO(8)$ are active. The naming reflects the fact that along this branch, generically 3 of the charges are set equal to one another, while the fourth is distinct. In this language, the ``+1" charge takes one from the 3+0 Q--black brane studied here to the 4Q  ($AdS$--RN) solution in which all four charges are set equal.

Fermion response from supergravity was studied in these backgrounds in \cite{DeWolfe:2014ifa}, and the authors described a fairly rich array of behaviors in the dual states of the ABJM theory. These include states with multiple Fermi surfaces, or even thick Fermi ``shells". Understanding  if or how these features are reflected in charge density correlation functions would be a logical and potentially illuminating continuation of the line of research we've advanced here.

Alternatively, from the perspective of the gauged SUGRA, one may wonder if there is a deeper symmetry at play which (approximately) relates branch points in the susceptibility to poles in the fermion Green's functions. In a supersymmetric background, bulk fermion and boson modes related by supersymmetry will share the same analytic structure in their dual correlation functions. A nice example of this appears in \cite{Bianchi:2000sm}. Although the 3Q--black brane preserves no supersymmetry, one can nevertheless ask whether it is possible that the supersymmetry is broken sufficiently softly such that an approximate relationship still holds in these backgrounds. Again, this question could be explored in more detail by moving along a branch of solutions in the $U(1)^4 \subset SO(8)$ truncation of the gauged SUGRA. Starting from the ``2+0" solution, which is supersymmetric, and turning on a small amount of ``+2" charge may allow one to gather some intuition for the effects of broken supersymmetry on the analytic 
structure of holographic correlation functions. We hope to report on this line of inquiry in a forthcoming sequel.

\acknowledgments{
We thank Mike Blake and especially Aristomenis Donos for collaboration on early incarnations of this work. We have also benefited immensely from collaboration on various related topics with Oliver DeWolfe, Jerome Gauntlett, and Steven Gubser, and thank O. DeWolfe for providing helpful comments on an advanced copy of this preprint. For illuminating and valuable discussion, we further thank Sebastian Fischetti, and in particular Blaise Gouteraux. 

The work of O.H. was supported by the Department of Energy under Grant No. DE-FG02-91-ER-40672, and by a Dissertation Completion Fellowship from the Graduate School at the University of Colorado Boulder. The work of C.R. was supported by the European Research Council under the European Union's Seventh Framework Programme (FP7/2007-2013), ERC Grant agreement ADG 339140. }

\appendix
\section{Finite Counterterms and the Alternate Quantization}\label{app:HRG}
The importance of a holographic renormalization procedure in the computation of many gauge theory quantitities via $AdS$/CFT is well known. Such a procedure is necessary to remove various near boundary divergences, and also plays a role in precision matches between gravity and gauge theory (recent examples include \cite{Freedman:2013ryh,Bobev:2013cja,Karch:2015kfa}). Comparatively, the role of {\it finite} boundary counterterms in holography has received much less attention. In part, this is because in many cases (including ``bottom-up" implementations of holography) there is no obvious method by which the coefficients of such terms can be unambiguously determined. At the same time, such terms have little physical significance in a broad class of holographic computations.

The situation changes somewhat in certain well known examples of top-down gauge/gravity duality. In these cases, the large amount of supersymmetry in the boundary gauge theory places strong constraints on the structure of finite counterterms, and this symmetry can sometimes be used to fix these terms completely. This procedure has recently been performed in detail in the $\mathcal{N}=8$ gauged SUGRA dual to the ABJM theory in the limit we study in the body of this work \cite{Freedman:2016yue}. 

Interestingly, finite boundary counterterms in this gauged SUGRA have profound consequences for the holographic dictionary. This is perhaps most evident for the scalars transforming as the ${\bf 35}_v$ of the $SO(8)$ gauge symmetry.  Consistency of the SUGRA theory requires that such modes obey an ``alternate" quantization, and in this case the addition of a finite counterterm to the on-shell boundary action can effectively redefine the identification of the source for the dual operator from the perspective of the gauge theory. 

To see this in more detail\footnote{We are indebted to O. DeWolfe and J. Gauntlett for discussions that greatly clarified these details.}, it is convenient to start from the renormalized action for a scalar in the ${\bf 35}_v$ which obeys alternate boundary conditions. Expanding (\ref{eq:Lag}) to quadratic order, the scalar sector is given by
\begin{equation}
S_{\phi} = \int\mathrm{d}^4x\sqrt{-g}\left( -\frac{1}{2}(\partial\phi)^2+\phi^2\right)
\end{equation}
plus the usual (infinite) alternate quantization boundary counterterms
\begin{equation}
S_{\mathrm{c.t.}} = \int\mathrm{d}^3 x\sqrt{-\gamma}\left(\frac{\phi^2}{2}+r\phi\phi' \right)
\end{equation}
where $'$ is a derivative with respect to $r$. The counterterm action is understood to be evaluated at some cutoff near the boundary which will eventually be taken to infinity. To these conventional terms, we now include the finite boundary counterterm 
\begin{equation}
S_{\mathrm{f.c.t.}}=\lambda\int\mathrm{d}^3 x\sqrt{-\gamma}\phi^3
\end{equation}
and seek boundary conditions on the bulk scalar such that the variational problem is well defined. 

Near the boundary the scalar falls off like
\begin{equation}
\phi(r\to\infty)\approx \frac{\alpha}{r}+\frac{\beta}{r^2}+\ldots
\end{equation}
and thus a variation of the renormalized on-shell action $\tilde{S}_{\mathrm{o.s.}} \equiv S_\phi^{\mathrm{o.s.}}+S_{\mathrm{c.t.}}+S_{\mathrm{f.c.t.}}$ leads to 
\begin{equation}\label{eq:bvar}
\delta\tilde{S}_{\mathrm{o.s.}}  = -\alpha\,\delta\beta+3\lambda\alpha^2\,\delta\alpha.
\end{equation}
Note that the conventional alternate boundary conditions, which identify the source for the scalar operator dual to $\phi$ as $\beta$ such that $\delta\beta = 0$, no longer extremize the action in the presence of the finite counterterm. Instead one can rewrite (\ref{eq:bvar}) as
\begin{equation}
\delta\tilde{S}_{\mathrm{o.s.}}  = -\alpha\,\delta\left(\beta -\frac{3}{2}\lambda\alpha^2 \right)
\end{equation}
which, upon choosing $\lambda = 1/3\sqrt{3}$, yields the quantization condition (\ref{eq:impQNT}).

\section{Boundary Conditions and Numerical Methods}
In order to compute the static susceptibility, we must solve the linearized fluctuation equations for $a_t$, and the modes that couple to $a_t$, while imposing the correct boundary conditions. The setup is very similar to \cite{Blake:2014lva}, with the added complication of the scalar field $\varphi$. We make the gauge choice $h_{\mu r} = a_r=0$. Inserting the ``background$+$fluctuation"  and plane wave ansatzes of (\ref{eq:bkgdPLUSfluct}) and (\ref{eq:planeWave}) into the bulk Einstein, Maxwell, and Klein-Gordon equations, we arrive at a set of ODEs that couple together the modes $\{a_t, h_{tt}, h_{yy}, \varphi \}$, as anticipated from the set of gauge invariant modes (\ref{eq:sZ2}). These equations are first order in $h_{tt}$ and second order in the other three modes. Hence there are seven constants of integration that must be fixed by boundary conditions.

We solve the equations exclusively in the non-zero temperature 3-charge black brane geometry. The IR thus has an event horizon, at which we impose regular boundary conditions:
\begin{align}
 a_t &\approx c_1 (r-r_H) + \mathcal{O}((r-r_H)^2) \\
 h_{yy} &\approx c_2 + \mathcal{O}(r-r_H) \\
 h_{tt} &\approx c_3 (r-r_H) + \mathcal{O}((r-r_H)^2) \\
 \varphi &\approx c_4 + \mathcal{O}(r-r_H)
\end{align}
The equations of motion fix \textit{e.g.} $c_3$ in terms of the other $c_i$. Hence we are left with three undetermined constants, which are to be fixed by UV boundary conditions.

When prescribing appropriate boundary conditions in the UV, extra care must be taken since we are working in terms of gauge dependent fluctuations. To compute the static susceptibility, one must ensure that the only non-normalizable fluctuation turned on belongs to $a_t$.

The scalar is quantized according to (\ref{eq:impQNT}), and thus normalizable linearized fluctuations of the scalar obey
\begin{equation}\label{eq:fctFLCT}
\frac{\delta\beta}{\sqrt{3}}-\frac{1}{3}\bar{\alpha}\,\delta\alpha = 0.
\end{equation}
In Fefferman-Graham coordinates, the fluctuation fall-offs in the UV ($r \rightarrow \infty$) are related to those of (\ref{eq:FOs}) by
\begin{align}
\varphi =&\, \frac{f_1}{\sr}+\frac{f_2+\frac{3Q}{4}f_1}{\sr^2}+\ldots\\
\tilde{h}_{tt} = & \,ht_0 + \ldots\\
\tilde{h}_{yy} = & \,hy_0 + \ldots\\
a_t = &\, a_0 + \frac{a_1}{\sr}+ \ldots
\end{align}
 Thus, to compute the static susceptibility it would naively seem as though one should seek solutions such that
\begin{equation}\label{eq:nbcs}
f_2+\frac{Q}{4}f_1 = ht_0 = hy_0 = 0 \qquad \mathrm{and} \qquad a_0 =1.
\end{equation}

However, since the earlier near-horizon analysis only provided us with three undetermined constants, it is clear that fixing these four boundary conditions would overconstrain the system. Instead, one is only permitted to enforce boundary conditions that are gauge equivalent to (\ref{eq:nbcs}). Turning to the near boundary behavior of the gauge invariant modes, 
\begin{align}
\mathcal{Z}_s=&\, \left(f_1+\frac{1}{4}\sqrt{3}Q\,hy_0\right)\frac{1}{\sr}+\left(\frac{3Q}{4}f_1+f_2+\frac{1}{8}\sqrt{3}Q^2\,hy_0 \right)\frac{1}{\sr^2}+\ldots\\
\mathcal{Z}_1 = & \,(ht_0 + hy_0)\,\sr^2+\ldots\\
\mathcal{Z}_2 = &\, a_0 -\left(\frac{(Q+r_H)\sqrt{Q(Q+r_H)}}{2}\,hy_0-a_1\right) \frac{1}{\sr}+ \ldots
\end{align}
one finds that the correct prescription is given by 
\begin{equation}\label{eq:ibcs}
f_2+\frac{Q}{4}f_1 = ht_0 +hy_0 = 0 \qquad \mathrm{and} \qquad a_0 =1.
\end{equation}
Any solution satisfying (\ref{eq:ibcs}) can be brought to the form (\ref{eq:nbcs}) by an appropriate coordinate transformation.

Having determined the correct boundary conditions for the IR and the UV, we employ numerical ``shooting" to integrate the fluctuations between the two. Starting from the IR near-horizon expansion, we organize the undetermined constants into a vector $\vec c \equiv (c_1,c_2,c_4)^T$, and set them to some arbitrary values, say $\vec c = (1,-1,1)^T$. Then, performing the numerical integration out to the UV boundary, we read off the values of the resulting sources---which we also collect into a vector $\vec J \equiv (f_2+\frac{Q}{4}f_1,ht_0 +hy_0,a_0)^T$. In general, this procedure will not lead to the desired UV boundary behavior $\vec J = (0,0,1)^T$. To remedy this, we use the fact that our fluctuation equations are linear. Repeating the numerical integration three times, with three linearly independent $\vec c$, allows us construct the linear map $\mathbf{T}$ from the set of IR data to the UV sources, defined by
\begin{equation}
 \mathbf{T} \vec c = \vec J \, .
\end{equation}
With $\mathbf{T}$ in hand, it is easy to find the vector $\vec c$ which corresponds to any desired $\vec J$, and to then compute the susceptibility using (\ref{eq:ssus}).

The matrix $\mathbf{T}$ is especially useful when searching for poles in the static susceptibility, which on the gravity side correspond to zero modes (ZMs) in the fluctuation spectrum. As explained in more detail in \cite{Kaminski:2009dh}, a ZM implies that $\mathbf{T}$ has a zero eigenvalue, and hence 
\begin{equation}
 \det \mathbf{T}(\vec k_{ZM}) = 0
\end{equation}
gives the condition for a pole in $\chi(k)$ at $\vec k=\vec k_{ZM}$. Combining this criterion with an efficient numerical root-finding algorithm (such as \textsl{Mathematica}'s \textbf{FindRoot}) provides a powerful method for locating ZMs and tracking their trajectory while changing e.g. the temperature.

\bibliography{oscBib}{}
\bibliographystyle{JHEP}

\end{document}